\documentclass[aps,pra,twocolumn,showpacs,amsmath,amssymb,preprintnumbers,superscriptaddress,10pt]{revtex4-1}

\usepackage{amsmath,amssymb}
\usepackage{bm}
\usepackage[latin1]{inputenc}
\usepackage{graphicx}
\graphicspath{{figures/}}
\usepackage{SIunits}
\usepackage{hyphenat}
\usepackage[bookmarks=false]{hyperref}
\usepackage{color}
\usepackage[capitalise]{cleveref}

\definecolor{pink}{RGB}{255,0,255}

\definecolor{ss_color}{rgb}{1,0,0}

\definecolor{darkorange}{RGB}{255,120,0}

\begin{document}

\title{Testing random-detector-efficiency countermeasure in a commercial system reveals a breakable unrealistic assumption}

\author{Anqi~Huang}
\email{angelhuang.hn@gmail.com}
\affiliation{Institute for Quantum Computing, University of Waterloo, Waterloo, ON, N2L~3G1 Canada}
\affiliation{\mbox{Department of Electrical and Computer Engineering, University of Waterloo, Waterloo, ON, N2L~3G1 Canada}}

\author{Shihan~Sajeed}
\affiliation{Institute for Quantum Computing, University of Waterloo, Waterloo, ON, N2L~3G1 Canada}
\affiliation{\mbox{Department of Electrical and Computer Engineering, University of Waterloo, Waterloo, ON, N2L~3G1 Canada}}

\author{Poompong~Chaiwongkhot}
\affiliation{Institute for Quantum Computing, University of Waterloo, Waterloo, ON, N2L~3G1 Canada}
\affiliation{Department of Physics and Astronomy, University of Waterloo, Waterloo, ON, N2L~3G1 Canada}

\author{Mathilde~Soucarros}
\affiliation{ID~Quantique SA, Chemin de la Marbrerie~3, 1227 Carouge, Geneva, Switzerland}

\author{Matthieu~Legr{\' e}}
\affiliation{ID~Quantique SA, Chemin de la Marbrerie~3, 1227 Carouge, Geneva, Switzerland}

\author{Vadim~Makarov}
\affiliation{Institute for Quantum Computing, University of Waterloo, Waterloo, ON, N2L~3G1 Canada}
\affiliation{Department of Physics and Astronomy, University of Waterloo, Waterloo, ON, N2L~3G1 Canada}
\affiliation{\mbox{Department of Electrical and Computer Engineering, University of Waterloo, Waterloo, ON, N2L~3G1 Canada}}


\begin{abstract}
In the last decade, efforts have been made to reconcile theoretical security with realistic imperfect implementations of quantum key distribution (QKD). Implementable countermeasures are proposed to patch the discovered loopholes. However, certain countermeasures are not as robust as would be expected. In this paper, we present a concrete example of ID Quantique's random-detector-efficiency countermeasure against detector blinding attacks.  As a third-party tester, we have found that the first industrial implementation of this countermeasure is effective against the original blinding attack, but not immune to a modified blinding attack. Then, we implement and test a later full version of this countermeasure containing a security proof [C.\ C.\ W.\ Lim {\it et al.,}\ IEEE J.\ Sel.\ Top.\ Quantum Electron.\ {\bf 21}, 6601305 (2015)]. We find that it is still vulnerable against the modified blinding attack, because an assumption about hardware characteristics on which the proof relies fails in practice. 
\end{abstract}

\maketitle

\section{Introduction}
\label{sec:introduction}

Currently, applied cryptography systems rely on the hardness of certain mathematical assumptions, which only provides computational security~\cite{naor2003, ETSI2015}. Once an eavesdropper has enough computing power, such as a quantum computer, the security of these classical encryption algorithms will be broken~\cite{bennett2000, shor1994}. However, quantum key distribution (QKD) allows two parties, Alice and Bob, to share a secret key based on the laws of quantum mechanics~\cite{bennett1984, ekert1991, gisin2002, scarani2009}. Because of no-cloning theorem \cite{wootters1982}, an eavesdropper with arbitrary computing power cannot copy the information sent by Alice without leaving any trace, which guarantees the unconditional security of communication~\cite{lo1999, shor2000, lutkenhaus2000, mayers2001, gottesman2004, renner2005}.

For this gradually maturing technology, practical QKD systems have been realised in laboratories~\cite{bennett1992, schmitt-manderbach2007, stucki2009, tang2015} and several companies have provided commercial QKD systems to general customers~\cite{QKDcompanies}. However, imperfect components used in the implementations lead to security issues that have attracted an increasing attention in the last decade~\cite{vakhitov2001, makarov2006, gisin2006, qi2007, zhao2008, lydersen2010a, sun2011, lydersen2011c,  jouguet2013, sajeed2015a}. Since increasing number of quantum attacks have been demonstrated, academic community is already aware of the security threat from practical loopholes. Therefore, the next step is to come up with loophole-free countermeasures. Importantly, the security of these countermeasures should be verified.

In this paper, an example of testing the security of an implemented countermeasure is given. We examine ID~Quantique's attempted countermeasure to earlier discovered bright-light detector control attacks~\cite{lydersen2010a, lydersen2010b, wiechers2011} that were demonstrated 6 years ago on ID~Quantique's and MagiQ Technologies' QKD products. The countermeasure is to randomly remove some detector gates to force the effective detection efficiency to zero during those slots~\cite{legre2010}. The idea is that when an eavesdropper is performing the blinding attack, she will produce click during these removed gates and thus get caught. This countermeasure has been implemented in a commercial system Clavis2 by two authors of this paper working at ID~Quantique (M.S.\ and M.L.), then provided as-is in a form of firmware update to the remaining four authors from the University of Waterloo who played the role of a third-party testing team. The authors from ID~Quantique did not participate in the test, however results of the test produced by the testing team were discussed by all authors and agreed upon.

The experimental results produced by the testing team show that although this countermeasure is effective against the original detector blinding attack~\cite{lydersen2010a}, it is no longer effective if the eavesdropper modifies her attack slightly. We note here that this countermeasure implemented by ID~Quantique is the simplest possible version of the original countermeasure proposal~\cite{legre2010}, and has already been criticised as unreliable in a later theoretical work~\cite{lim2015}. Hence, the testing team has gone further ahead and manually implemented a full version of the countermeasure using two non-zero detection efficiency levels~\cite{legre2010, lim2015}, and tested it. Our testing shows that even the full countermeasure is vulnerable to the modified blinding attack. Specifically, we experimentally disprove an assumption that Bob's detection probability under blinding attack cannot be proportional to his single-photon detection efficiency, on which the theoretical analysis in Ref.~\onlinecite{lim2015} relies.

The paper is organized as follows. \cref{sec:random-gate-removal-countermeasure} reviews a hacking-and-patching timeline of ID~Quantique's Clavis2 QKD system and introduces the countermeasure. In~\cref{sec:hacking-the-countermeasure}, testing results of ID~Quantique's first countermeasure implementation are reported and our modified blinding attack is introduced. \Cref{theory} theoretically analyses conditions of a successful attack and shows that the modified blinding attack satisfies them. Moreover, in~\cref{sec:discussion-of-full-countermeasure}, based on certain assumptions about a future implementation of the full countermeasure~\cite{lim2015}, we demonstrate two possible methods to hack this full version implementation. We discuss the practicality of our attacks against installed commercial QKD lines in \cref{sec:attack-black-box} and conclude in \cref{sec:conclusion}.

\section{From loophole discovery to countermeasure implementation}
\label{sec:random-gate-removal-countermeasure}

\begin{figure}
	\includegraphics[width=\columnwidth]{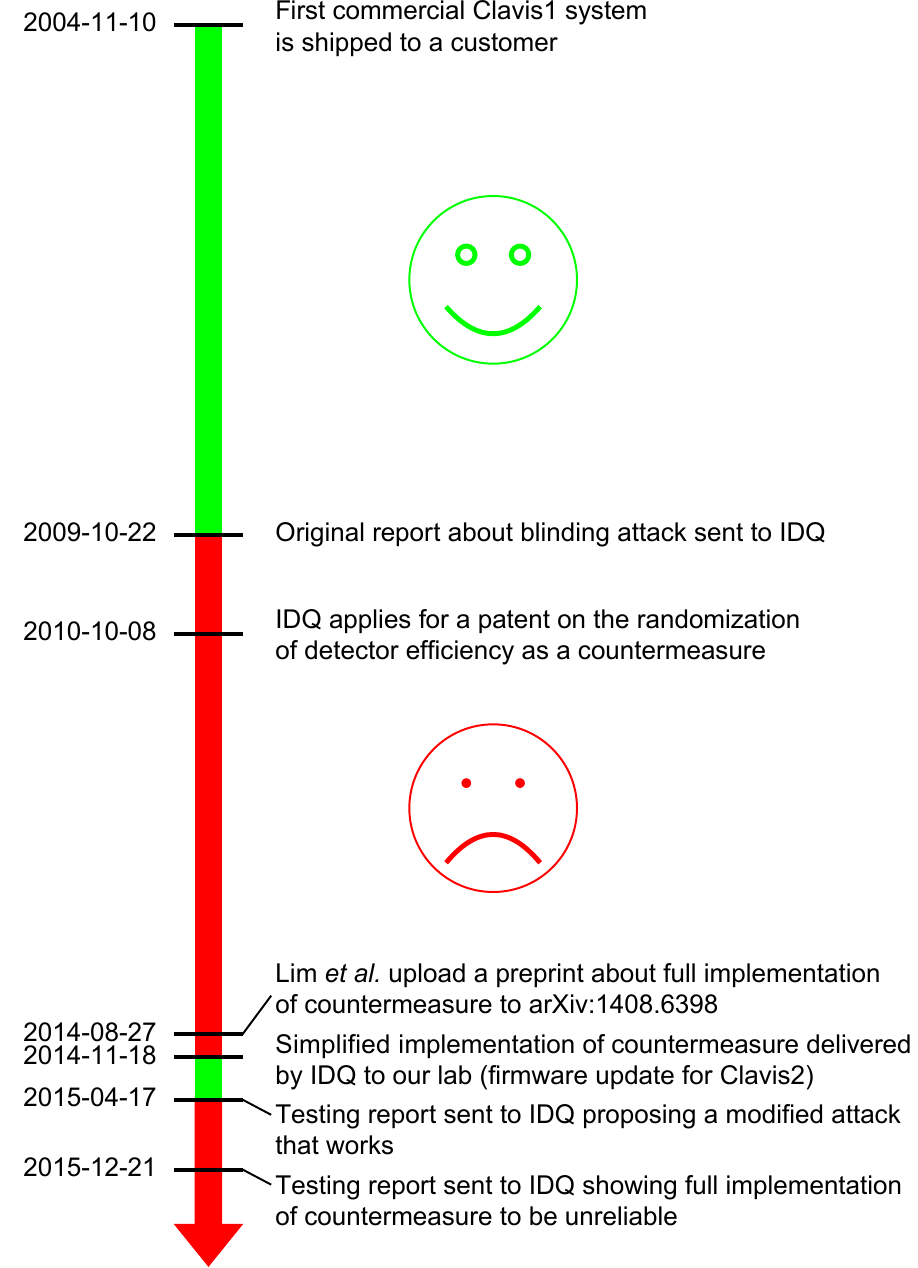}
	\caption{Timeline of hacking-countermeasure-hacking for the bright-light detector control class of attacks.}
	\label{fig:timeline}
\end{figure}

In 2009, the vulnerability of the commercial QKD system Clavis2 \cite{idqclavis2specs} to detector blinding attacks was identified and a confidential report was submitted to ID~Quantique (the work was published shortly afterwards \cite{lydersen2010a}). After this, ID~Quantique has been trying to figure out an experimental countermeasure against these attacks. The timeline of this security problem is shown in \cref{fig:timeline}. In 2010, ID~Quantique proposed a countermeasure that randomizes the efficiency of a gated avalanche photodiode~(APD) by randomly choosing one out of two different gate voltages, and filed this idea for a patent~\cite{legre2010}. In this way, an eavesdropper Eve does not know the exact efficiency of Bob in every gated slot and thus cannot maintain his detection statistics. At the sifting phase, if the observed detection rates differ from the expected values, Alice and Bob would be aware of Eve's presence and discard their raw keys.

In 2014, Lim \textit{et al.}\ proposed a specific protocol to realize this countermeasure~\cite{lim2015}, which analyses the security mathematically for blinding attacks that obey a certain assumption on their behavior. In the protocol, Bob randomly applies two non-zero detection efficiencies $\eta_\text{1} > \eta_\text{2} > 0$, and measures detection rates $R_\text{1}$ and $R_\text{2}$ conditioned on these efficiencies. The effect of detector blinding attack is accounted via the factor $\left(\eta_\text{1}R_\text{2}-\eta_\text{2}R_\text{1}\right)/\left(\eta_\text{1}-\eta_\text{2}\right)$. Without the blinding attack, the detection rate is proportional to the efficiency, making this factor zero. The analysis makes a crucial assumption that the detection rate under blinding attack $R_\text{1} = R_\text{2}$, i.e.,\ it will be independent of Bob's choice of $\eta_\text{1,2}$. Then, under attack the factor will be greater than zero, and reduces the secure key rate. This solution intends to introduce an information gap between Eve and Bob, for Eve has no information about Bob's random efficiency choice.

Later in 2014, ID~Quantique implemented the countermeasure as a firmware patch. The hardware in Clavis2 is not capable of generating two nonzero efficiency levels that switch randomly between adjacent detector gates. As a result, implementation is in a simple form by suppressing gates randomly with $2\%$ probability. The suppressed gates represent zero efficiency $\eta_2 = 0$, while the rest of the gates represent calibrated efficiency $\eta_1 =\eta$.  Ideally, in the updated system, there should be no click in the absence of the gate. In practice, transient electromagnetic interference may extremely infrequently lead to a click without a gate. Therefore, an alarm counter is used with the system lifetime limit of 15 clicks in the absence of the gate. If this limit is reached, it triggers the firmware to brick the system and require factory maintenance. This implementation assumes that under blinding attack \cite{lydersen2010a}, click probability should not depend on the gate voltage and the attack should therefore cause clicks at the slots of gate absence.

\section{testing the countermeasure}
\label{sec:hacking-the-countermeasure}

\begin{figure}
	\includegraphics[width=\columnwidth]{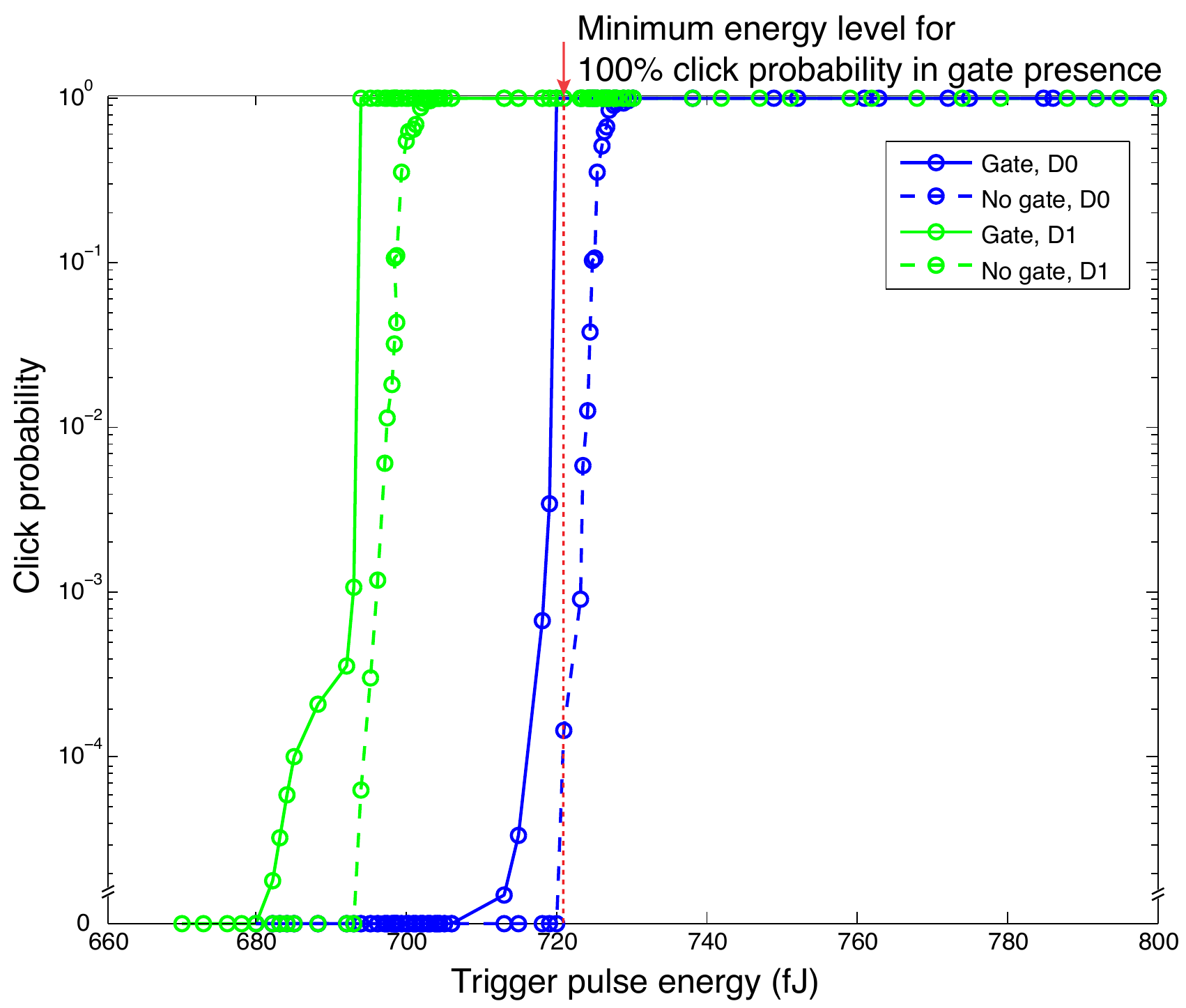}
	\caption{Click probability under original blinding attack~\cite{lydersen2010a} versus energy of trigger pulse. The blinding power is $1.08~\milli\watt$, as the same as the power used in the published original attack~\cite{lydersen2010a}. The timing of trigger pulse is $0.7~\nano\second$ long, $3~\nano\second$ after the centre of the gate signal, which should roughly reproduce the original attack~\cite{lydersen2010a}.}
	\label{fig:originalattack}
\end{figure}   

In this section, we demonstrate that the countermeasure presently implemented by ID Quantique is effective against the original blinding attack~\cite{lydersen2010a}, but not sufficient against the general class of attacks attempting to take control of Bob's single-photon detectors.

Let us briefly remind the reader how Clavis2 and the original blinding attack against it work. Clavis2 is a bidirectional phase-encoding QKD system~ \cite{stucki2002,idqclavis2specs}. After Bob sends multi-photon bright pulses to Alice, Alice randomly modulates one of the four BB84 phase states~\cite{bennett1984}, attenuates the pulses and sends them back to Bob. Bob randomly chooses one out of two measurement bases. Interference happens between pulses from longer and shorter paths of an interferometer at Bob's side, and the outcomes of interference depend on the phase difference between Alice's and Bob's modulation~\cite{muller1997}. However, Eve is able to control the outcomes by the following strategy. She shines bright light to blind the detectors, and then intercepts Alice's states~\cite{lydersen2010a}. According to Eve's interception results, she re-sends faked states by multi-photon pulses to Bob's blinded detectors. If Bob chooses the same measurement basis as Eve's, the pulses interfere at Bob's interferometer, so that all power of the pulse goes to one detector to trigger a click. If the measurement bases chosen by Bob and Eve are mismatched, there is no interference, and the power of the pulse is split equally between Bob's two detectors. In this case, neither detector clicks. In this attack, Eve can fully control Bob's detectors and obtain the whole key tracelessly~\cite{lydersen2010a}.

For the original blinding attack, Eve sends bright-light continuous-wave (c.w.)\ laser light to blind Bob's detectors. Then a trigger pulse is sent slightly after the gate to make a click. We repeat this attack for improved Clavis2 system and test the amount of energy to trigger a click which is shown in~\cref{fig:originalattack}. From~\cref{fig:originalattack}, we can see the trigger pulse energy for gate presence (solid curves) is lower than that for gate absence (dashed curves), because minute electrical fluctuations of APD voltage following the gate signal lower the click threshold slightly.

However, if Eve tries to trigger a click with $100\%$ probability when the gate is applied, this amount of trigger pulse energy (marked by a dotted vertical line in~\cref{fig:originalattack}) also might trigger a click with non-zero probability when the gate is suppressed, which is monitored and results in an alarm. Therefore, Eve cannot hack the system with full controllability. To avoid clicks in slots of gate suppression, Eve could in theory decrease the level of trigger pulse energy to trigger a click sometimes with gate presence, but never with gate absence. This also satisfies a necessary condition of a successful attack which we will discuss in~\cref{theory} later. Unfortunately, in practice, our testing result shows the amount of trigger pulse energy required to trigger D0 without the gate is about $710~\femto\joule$, which is only $1.5\%$ less than the amount of energy for $100\%$ click ($720~\femto\joule$) when the gate is present. The $1.5\%$ difference of these two energy levels is likely not big enough to achieve a reliable attack operation that avoids triggering the countermeasure. Also, D1 will always trigger at these energy levels, revealing the attack. Eve could target D1 using a slightly lower energy level, but the relative precision required is similar there. Routine fluctuations of temperature and other equipment parameters may lead to some instability of these trigger pulse energy levels, causing a risk for Eve to trigger a few clicks in the gate absence and brick the system being attacked. From this point of view, we think this first implementation of countermeasure is effective against the original blinding attack.

\begin{figure}
	\includegraphics[width=\columnwidth]{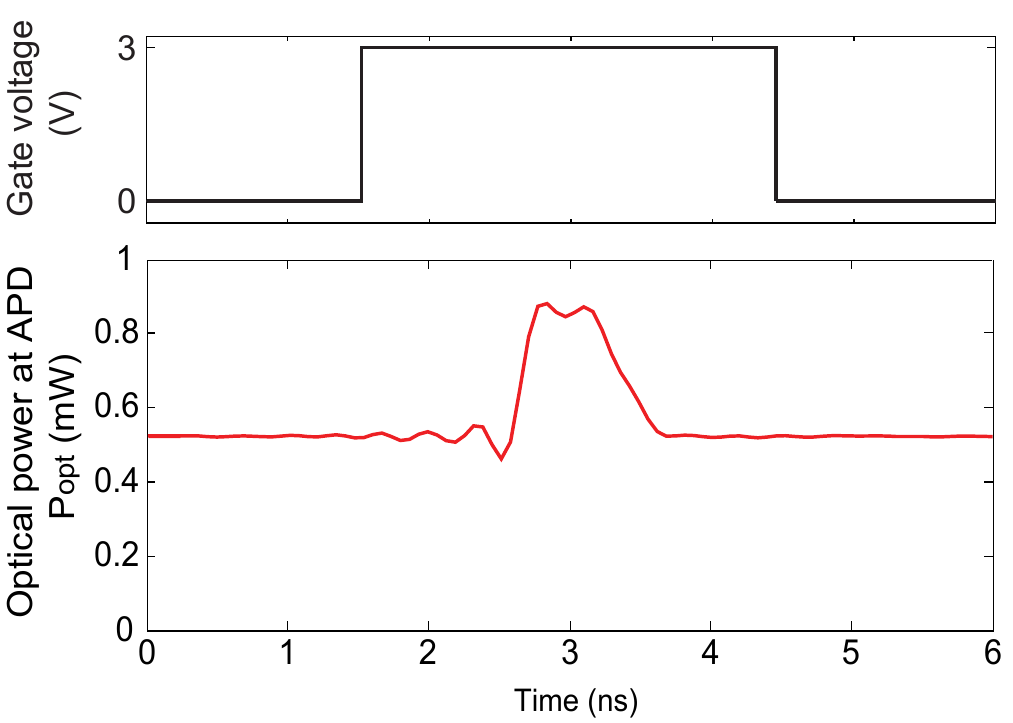}
	\caption{Idealized APD gate signal and real oscillogram of optical trigger pulse. Relative time between the gate voltage transitions and the optical pulse is approximate. The c.w. signal is generated by a $1536~\nano\meter$ laser diode; the trigger pulse signal is obtained by modulating pump current of a separate $1551~\nano\meter$ laser diode, using an electrical pulse generator \cite{ lydersen2010a}.}
	\label{fig:triggersingal}
\end{figure}   

\begin{figure*}
	\includegraphics[width=\textwidth]{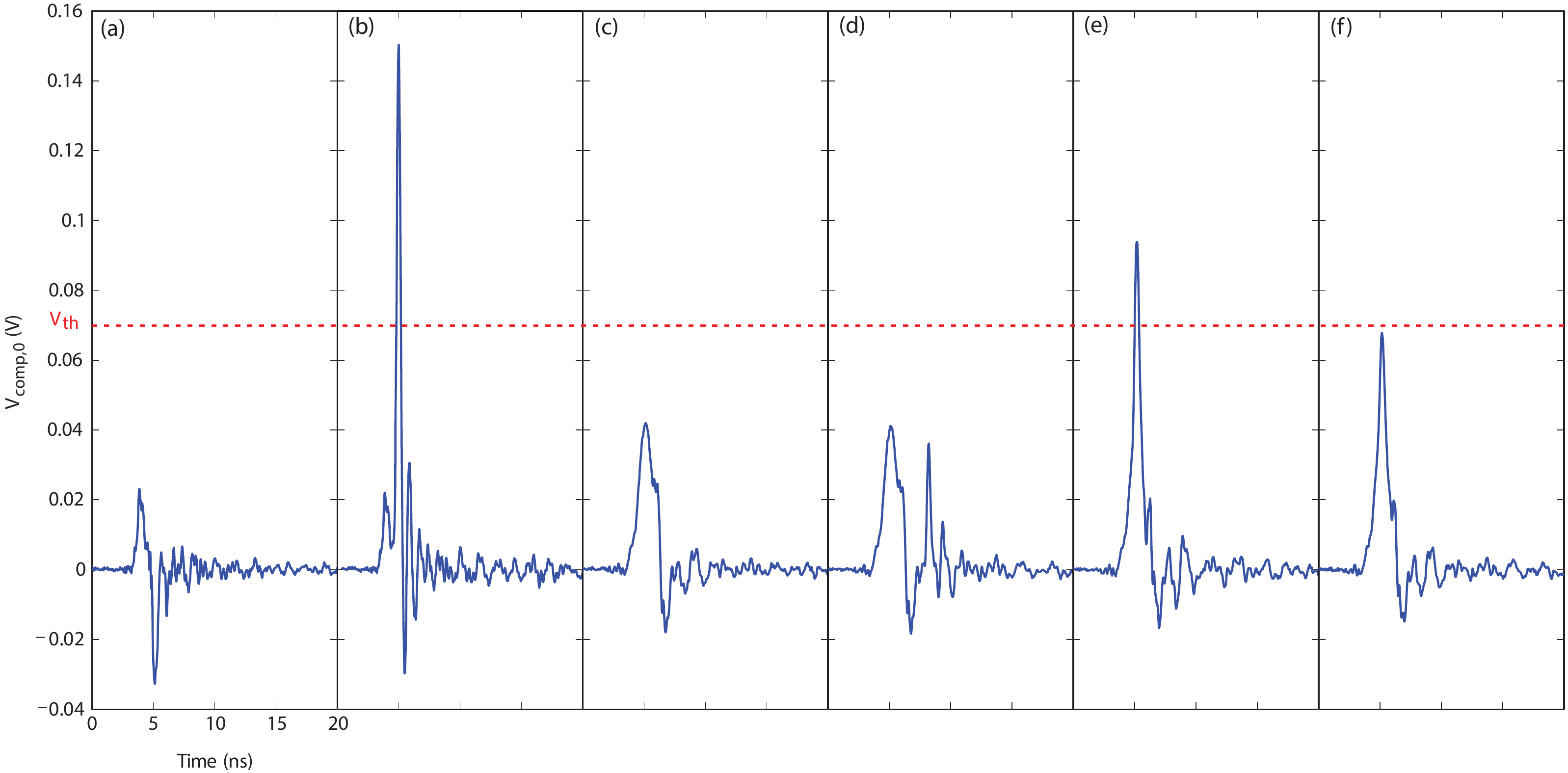}
	\caption{Oscillograms at comparator input in the detector circuit, proportional to APD current. (a)~Geiger mode. The small positive and negative pulses are due to gate signal leakage through the APD capacitance of~$\sim 1~\pico\farad$. (b)~Geiger mode, single-photon avalanche.  (c--f)~The detector is blinded with $0.56~\milli\watt$ c.w.\ illumination, with (c)~no trigger pulse applied, (d)~$0.32~\pico\joule$ trigger pulse applied $5~\nano\second$ after the gate, (e)~$0.32~\pico\joule$ trigger pulse applied in the gate, and (f)~$0.16~\pico\joule$ trigger pulse applied in the gate.}
	\label{fig:gatetriggersingal}
\end{figure*} 

We can slightly modify our blinding attack to break the security of this countermeasure. Similarly to the original blinding attack, Bob's detectors are blinded by a bright-light laser first. Then, instead of sending a trigger pulse slightly after the gate as in the original attacks~\cite{lydersen2010a}, we send a $0.7~\nano\second$ long trigger pulse on top of the c.w.\ illumination {\it during the detector gate}, as shown in \cref{fig:triggersingal}. This trigger pulse produces a click in one of Bob's two detectors only if Bob applies the gate and his basis choice matches that of Eve; otherwise there is no click.

To explain why this modified attack succeeds, let us remind the reader the normal operation of an avalanche photodiode (APD). The detectors in Clavis2 are gated APDs. When the gate signal is applied, the voltage across the APD $V_\text{APD}$ is greater than its breakdown voltage $V_\text{br}$. If a single photon comes during the gated time, an avalanche happens and causes a large current. This current is converted into a voltage by the detector electronic circuit. If the peak voltage is larger than a threshold $V_\text{th} = 70~\milli\volt$, the detector registers a photon detection (a `click'). \cref{fig:gatetriggersingal}(a) and (b) show the cases of no photon coming and a photon introducing an avalanche. \Cref{sec:background} explains more details of the detector operation principle and the blinding attack.

A bright laser is able to blind the APDs. Under c.w.\ illumination, the APD produces constant photocurrent that overloads the high-voltage supply and lowers $V_\text{APD}$. Then, even when the gate signal is applied, $V_\text{APD}$ does not exceed $V_\text{br}$ and the APD remains in the linear mode as a classical photodetector that is no longer sensitive to single photons. This means the detectors become blinded.

Under the blinding attack, \cref{fig:gatetriggersingal}(c--e) shows the detector voltages in different cases: when no trigger pulse is applied and when the trigger pulse is applied either after or in the gate. Since in the linear mode the gain factor of secondary electron-hole pairs generation in the APD depends on the voltage across it, the $3~\volt$ gate applied to the APD increases the gain factor. This larger gain during the gated time assists the APD in generating a larger photocurrent than the photocurrent outside the gate. Therefore the gate signal causes a positive pulse as shown in \cref{fig:gatetriggersingal}(c). The trigger pulse applied after the gate produces a second pulse, but the peak voltages of neither pulses exceed $V_\text{th}$ [\cref{fig:gatetriggersingal}(d)]. However, when the trigger pulse is shifted inside the gate, the two pulse amplitudes add up, reach $V_\text{th}$ and produce a detector click [\cref{fig:gatetriggersingal}(e)]. If Bob chooses a different measurement basis than Eve, only half of the trigger pulse energy arrives at each detector~\cite{lydersen2010a}. In this case,  the peak voltage does not reach $V_\text{th}$ [\cref{fig:gatetriggersingal}(f)]. Overall, only when the trigger pulse is applied during the gate time and Bob chooses the same basis as Eve, the detector under the blinding attack clicks. As a result, Eve can control Bob's detectors to make Bob obtain the same measurement result as her, and does not introduce extra errors~\cite{lydersen2010a}.

\begin{figure}
	\includegraphics{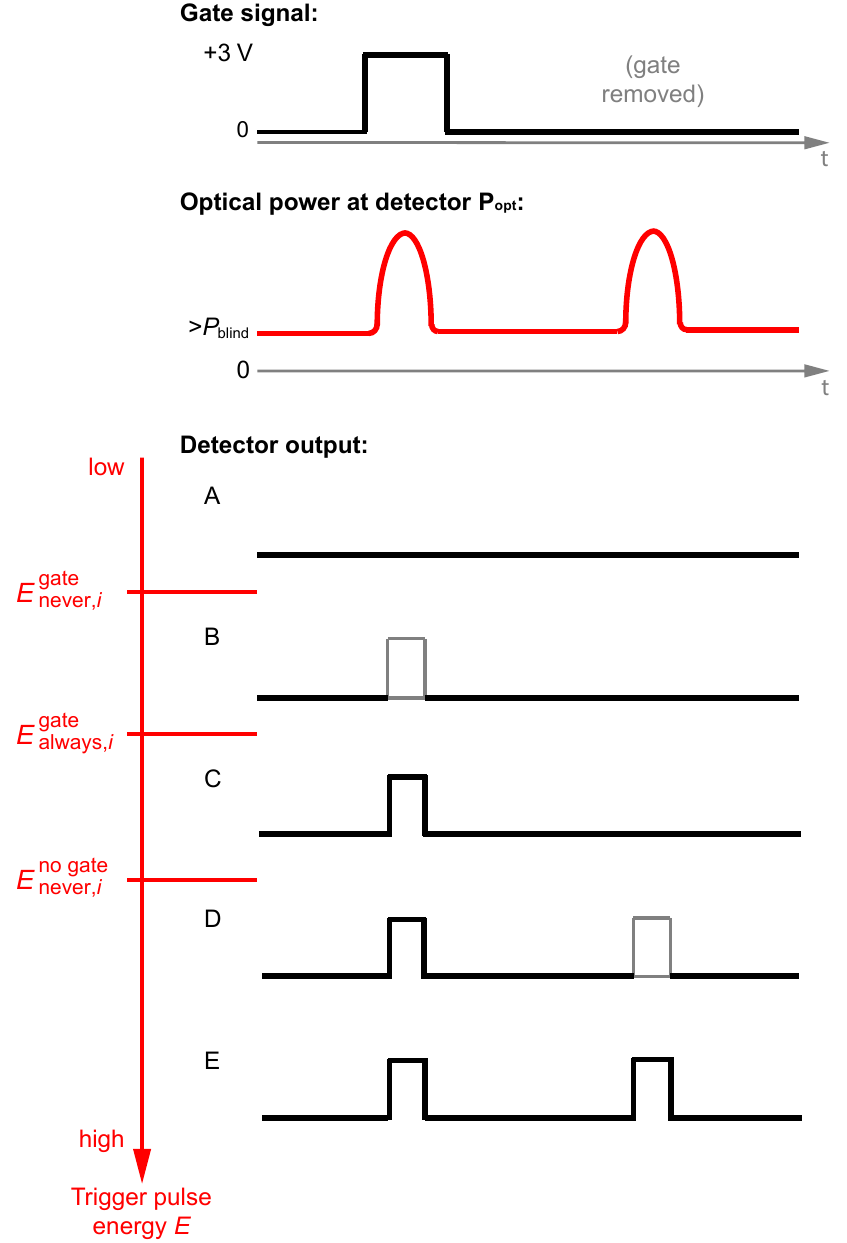}
	\caption{Output of a blinded detector in Clavis2 under control of trigger pulses of different energy. The top graph shows a gate applied at the first slot, but suppressed at the second slot. However, an optical trigger pulse is sent to the detector in both slots. Graphs A--E show detector output versus trigger pulse energy $E$. In graph A, the energy is insufficient to produce a click. As the energy is increased above $E^\text{gate}_{\text{never,}i}$, clicks intermittently appear in the presence of the gate, as shown in graph~B. At the energy level above $E^\text{gate}_{\text{always,}i}$, the gate always has a click, as shown in graph~C. However, there is never a click when there is no gate. At a higher energy level above $E^\text{no\,gate}_{\text{never,}i}$, clicks in the gate absence appear intermittently (graph~D) or always (graph~E). }
	\label{fig:cases}
\end{figure} 

Contrary to most of previously demonstrated attacks attempting to take control of single-photon detectors~\cite{lydersen2010a, lydersen2010b, lydersen2011c}, in the present demonstration the timing of the trigger pulse has to be aligned with the gate. Besides timing alignment, another important factor of the attack is the trigger pulse energy $E$. To test the effect of different trigger pulse energy, we gradually increase it and observe the detection outcomes. \Cref{fig:cases} shows schematically in which order clicks appear in Clavis2 as $E$ is increased. We observe three thresholds.

\begin{itemize}
	\item If $E \leq E^\text{gate}_{\text{never,}i}$ (where $i \in \left\{0,1\right\}$ is detector number), the detector never clicks when the gate is applied.
	
	\item If $E \geq E^\text{gate}_{\text{always,}i}$, the detector always clicks when the gate is applied.
	
	\item If $E \leq E^\text{no\,gate}_{\text{never,}i}$, the detector never clicks when the gate is suppressed.
\end{itemize}

\begin{figure}
	\includegraphics[width=\columnwidth]{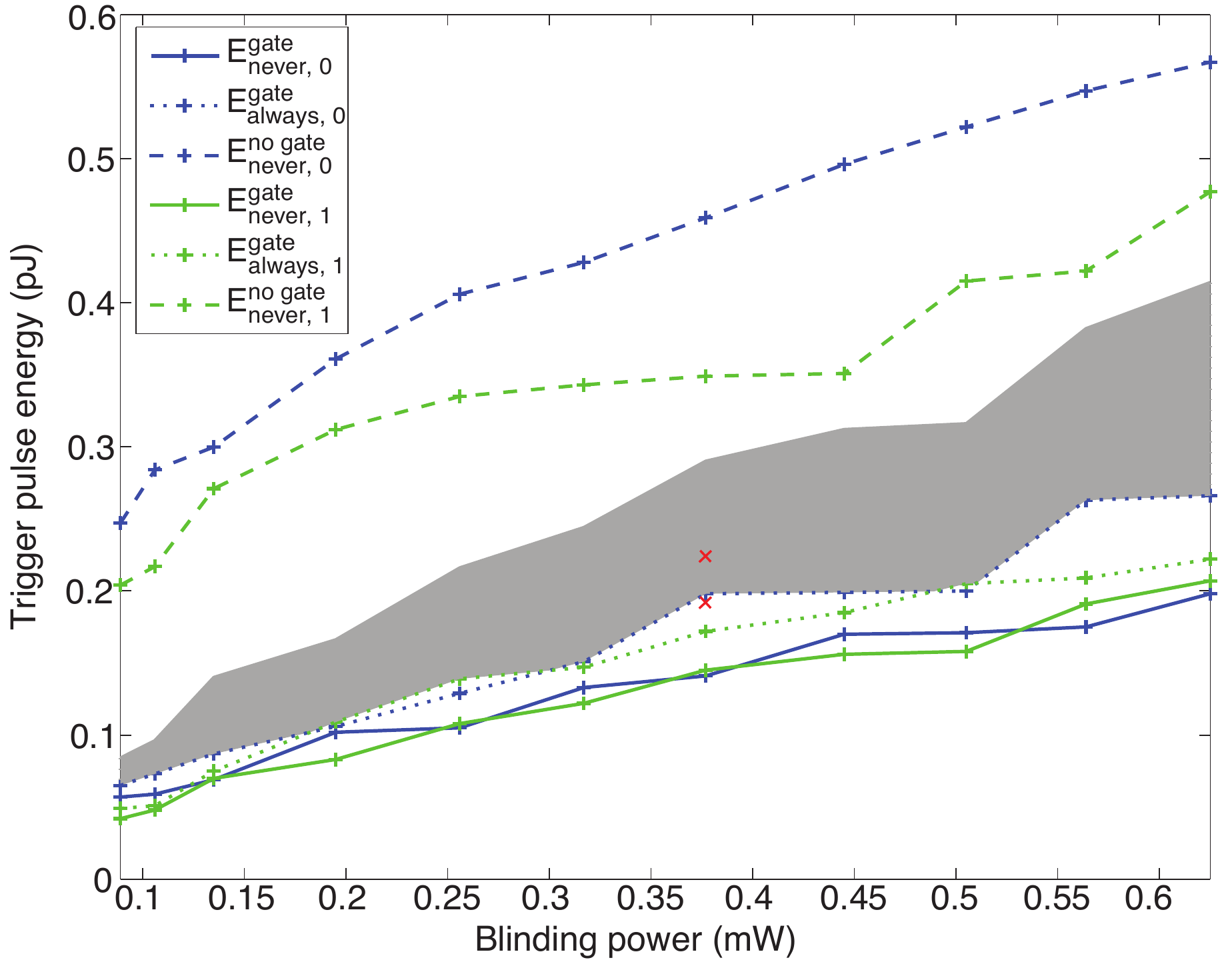}
	\caption{Energy thresholds of trigger pulse versus c.w.\ blinding power. Shaded area shows the range of trigger pulse energies of the perfect attack.}
	\label{fig:triggerresult}
\end{figure}

\Cref{fig:triggerresult} shows these detection thresholds measured for a range of c.w.\ blinding powers. All the thresholds rise with the blinding power, because higher blinding power leads to a larger photocurrent and lower $V_\text{APD}$. The decreased $V_\text{APD}$ leads to smaller gain and thus lower sensitivity to the trigger pulse. (\Cref{sec:understanding-apd-operation} contains a more detailed investigation of the processes inside the detector.) As can be seen, for any given blinding power, $E^\text{no\,gate}_{\text{never,}i}$ is much higher than the other click thresholds. This easily allows the original detector control attack~\cite{lydersen2010a} to proceed undetected by the countermeasure. A more formal analysis will be stated in the next section.

\section{Conditions of a successful attack}
\label{theory}

Experimental result of the previous section shows that the attack of Ref.\ \onlinecite{lydersen2010a} is possible in Clavis2. However, general conditions for a successful attack should be analysed theoretically. In this section, we first consider \textit{strong conditions} for a perfect attack, in which Eve induces a click in Bob with $100\%$ probability if their bases match and the gate is applied, and $0\%$ probability otherwise. These conditions are definitely sufficient for a successful attack~\cite{lydersen2010a}. However, as we remark later in this section, even if these strong conditions are not satisfied, an attack may still be possible.

\textbf{Strong conditions.} If the detection outcome varies as~\cref{fig:cases} with the increase of trigger pulse energy, the order of the three thresholds is:
\begin{equation}
\label{eq:strong-energy-thresholds}
E^\text{no\,gate}_{\text{never,}i} > E^\text{gate}_{\text{always,}i} > E^\text{gate}_{\text{never,}i}.
\end{equation}

If Eve and Bob select opposite bases, half of the energy of trigger pulse goes to each Bob's detector. In this case, none of the detectors should click despite the gate presence. This is achieved if \cite{lydersen2010a}
\begin{equation}
\label{eq:strong-basis-match-micmatch}
\frac{1}{2} \max_i\left\{E^\text{gate}_{\text{always,}i}\right\} < \left( \min_i\left\{E^\text{gate}_{\text{never,}i}\right\} \right).
\end{equation}
The random gate suppression imposes additional conditions. In case of basis mismatch, half of the trigger pulse energy is arriving at each detector. It should induce a click in neither detector when the gate signal is absent. For the target detector \textit{i}, there is no click once~\cref{eq:strong-energy-thresholds} is satisfied. For the other detector \textit{i}$\oplus 1$, no click is achieved when half of the trigger pulse energy is still lower than the detection threshold in the no-gate case. That is,
\begin{equation}
\label{eq:strong-nogate-basis-micmatch}
\frac{1}{2} E^\text{gate}_{\text{always,}i} < E^\text{no\,gate}_{\text{never,}i \oplus 1}.
\end{equation}
If the bases match, we need to make sure there is no click when the gate is suppressed, but always a click in the expected detector in the gate presence. This is achieved if $E^\text{gate}_{\text{always,}i} < E^\text{no\,gate}_{\text{never,}i}$, which is already included in inequality \labelcref{eq:strong-energy-thresholds}. Although inequality \labelcref{eq:strong-nogate-basis-micmatch} has a physical meaning, it mathematically follows from inequalities \labelcref{eq:strong-energy-thresholds,eq:strong-basis-match-micmatch}. Thus satisfying inequalities \labelcref{eq:strong-energy-thresholds,eq:strong-basis-match-micmatch} represents the strong attack conditions and guarantees the same performance as in Ref.\ \onlinecite{lydersen2010a}. The shaded area in~\cref{fig:triggerresult} indicates a range of the trigger pulse energies Eve can apply for the perfect attack. The range is sufficiently wide to allow for a robust implementation, only requiring Eve to set correct energy with about $\pm 15\%$ precision.

\textbf{Necessary condition.} An attack may still be possible even if Eve's trigger pulse does not always cause a click in Bob when their bases match, and/or sometimes causes a click when their bases do not match \cite{lydersen2011b}. The latter introduces some additional QBER but as long as it's below the protocol abort threshold, Alice and Bob may still produce key. The random gate removal countermeasure imposes the condition
\begin{equation}
\label{eq:necessary-energy-thresholds}
E^\text{no\,gate}_{\text{never,}i} > E^\text{gate}_{\text{never,}i},
\end{equation}
which means Eve should be able to at least sometimes cause a click in the gate while never causing a click without the gate (lest the alarm counter is increased). This is a necessary condition for an attack. As the present paper details, there are strong engineering reasons why this condition is likely to be satisfied in a detector. Additional conditions will depend on exact system characteristics \cite{lydersen2011b}.

\section{Will a full implementation of the countermeasure be robust?}
\label{sec:discussion-of-full-countermeasure}

We have proved so far that the current countermeasure with gate suppression cannot defeat the detector blinding attack. However, the paper of Lim {\it et al.}~\cite{lim2015} claims that the full version of countermeasure with two non-zero detection efficiencies is effective against a large class of detector side-channel attacks including the blinding attack~\cite{lydersen2010a}. Even though this full countermeasure has not been implemented by ID Quantique, we have tested some properties of the detectors in Clavis2 to show two possible methods to hack the full countermeasure, based on certain assumptions about a future implementation.

Bob could choose randomly between $P/2$ and $P$ detection efficiency by changing either gate voltage amplitude $V_\text{gate}$ or high-voltage supply $V_\text{bias}$~\cite{lim2015}. Since in Clavis2 hardware $V_\text{gate}$ is fixed (see \cref{sec:background}), we assume an engineer will change $V_\text{bias}$ to achieve different non-zero detection efficiencies. To achieve half of original detection efficiency, we lower $V_\text{bias}$ manually. When $V_\text{bias,0}$ of D0 drops from $-55.26~\volt$ to $-54.86~\volt$, the detection efficiency $P_0$ reduces from $22.6\%$ to $12.8\%$. Similarly, we decrease $V_\text{bias,1}$ of D1 from $-54.70~\volt$ to $-54.40~\volt$, leading to the detection efficiency $P_1$ reduction from $18.9\%$ to $9.7\%$. After that, we test Eve's controllability of these two detectors.

\begin{figure}
	\includegraphics[width=\columnwidth]{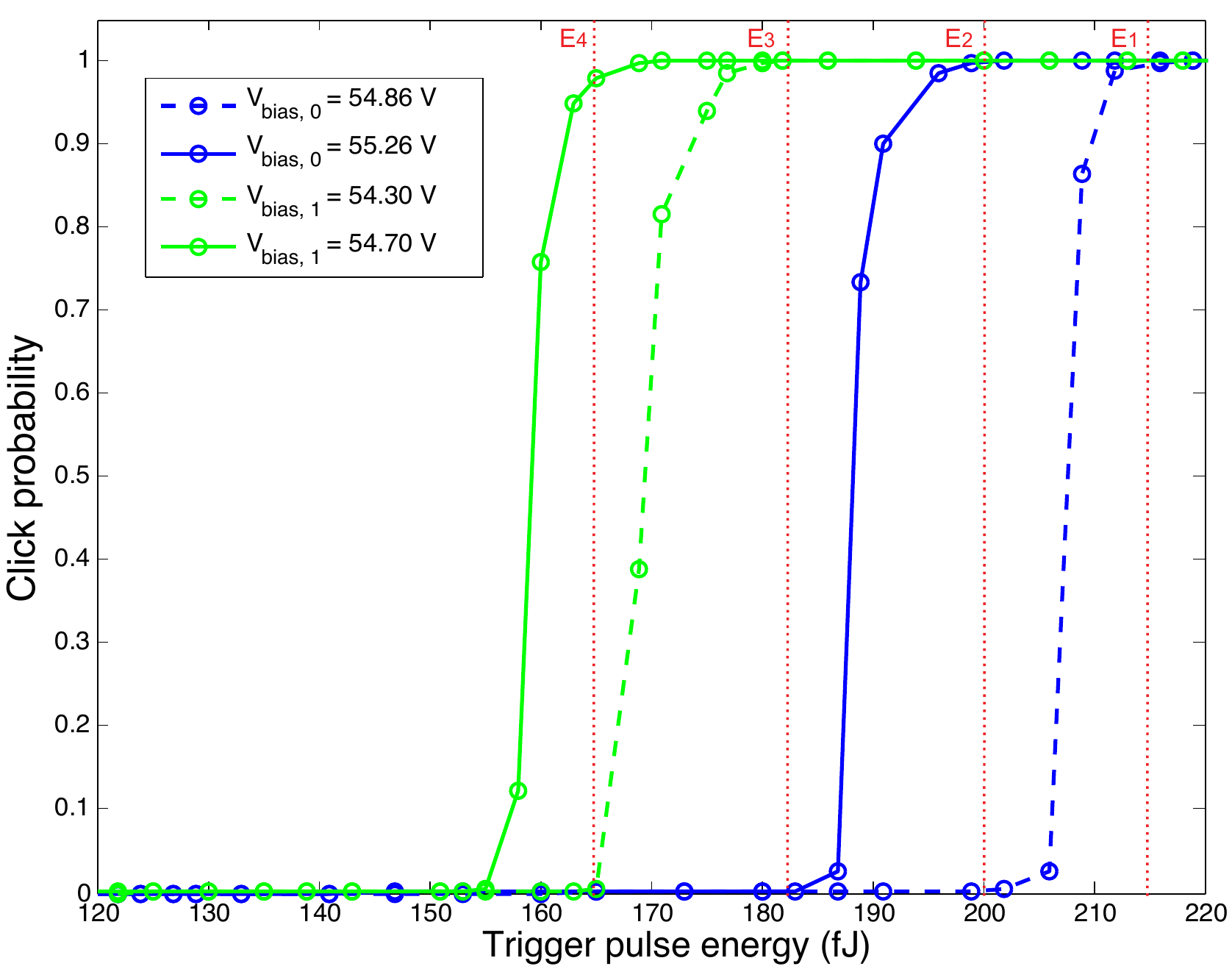}
	\caption{Click probabilities under blinding attack versus energy of trigger pulse. Solid curves show the energy of trigger pulse for original $V_\text{bias}$, while dashed curves for reduced $V_\text{bias}$ lowering photon detection efficiency by about a factor of 2. The blinding power is $0.38~\milli\watt$ and the timing of trigger pulse is aligned in the middle of the gate by minimizing its energy required to make a click.}
	\label{fig:detection_efficiency_energy}
\end{figure}

First, we blind the detectors and then measure the relation between the energy of trigger pulse and probability to cause a click. The position of trigger pulse is fixed in the middle of gate signal. \Cref{fig:detection_efficiency_energy} shows the testing result which indicates there is a transition range between $0\%$ and $100\%$ click probability.

From the measurement result, Eve can randomly select different levels of trigger pulse energy (shown as dotted lines in~\cref{fig:detection_efficiency_energy}) to attack the full version of countermeasure. As we know, only when Bob chooses the same measurement basis as Eve, all the energy of trigger pulse arrives targeted detector and achieves a click. For target D0, if trigger pulse energy $E_1$ is chosen, D0 always clicks, while at $E_2$, the detector only clicks if higher $V_\text{bias}$ is applied. When $E_1$ and $E_2$ are chosen randomly with the same probability $P_0/2$, the detection probability for higher $V_\text{bias}$ is $P_0$ and the detection probability for lower $V_\text{bias}$ is only $P_0/2$. Therefore, the attack reproduces correct detection probabilities as the protocol requires. Similarly, for target D1, Eve can choose $E_3$ to trigger click always and choose $E_4$ to get a click only if higher $V_\text{bias}$ is applies. This reproduces correct detection probabilities, $P_1/2$ and $P_1$. At the same time, $E_1$ and $E_3$ remain safely below $E_\text{never,0,1}^\text{no gate}$ shown in~\cref{fig:triggerresult}, so clicks are never produced in the absence of the gate and alarm is not triggered. This allows Eve to hack the countermeasure tracelessly.

\begin{figure}
	\includegraphics[width=\columnwidth]{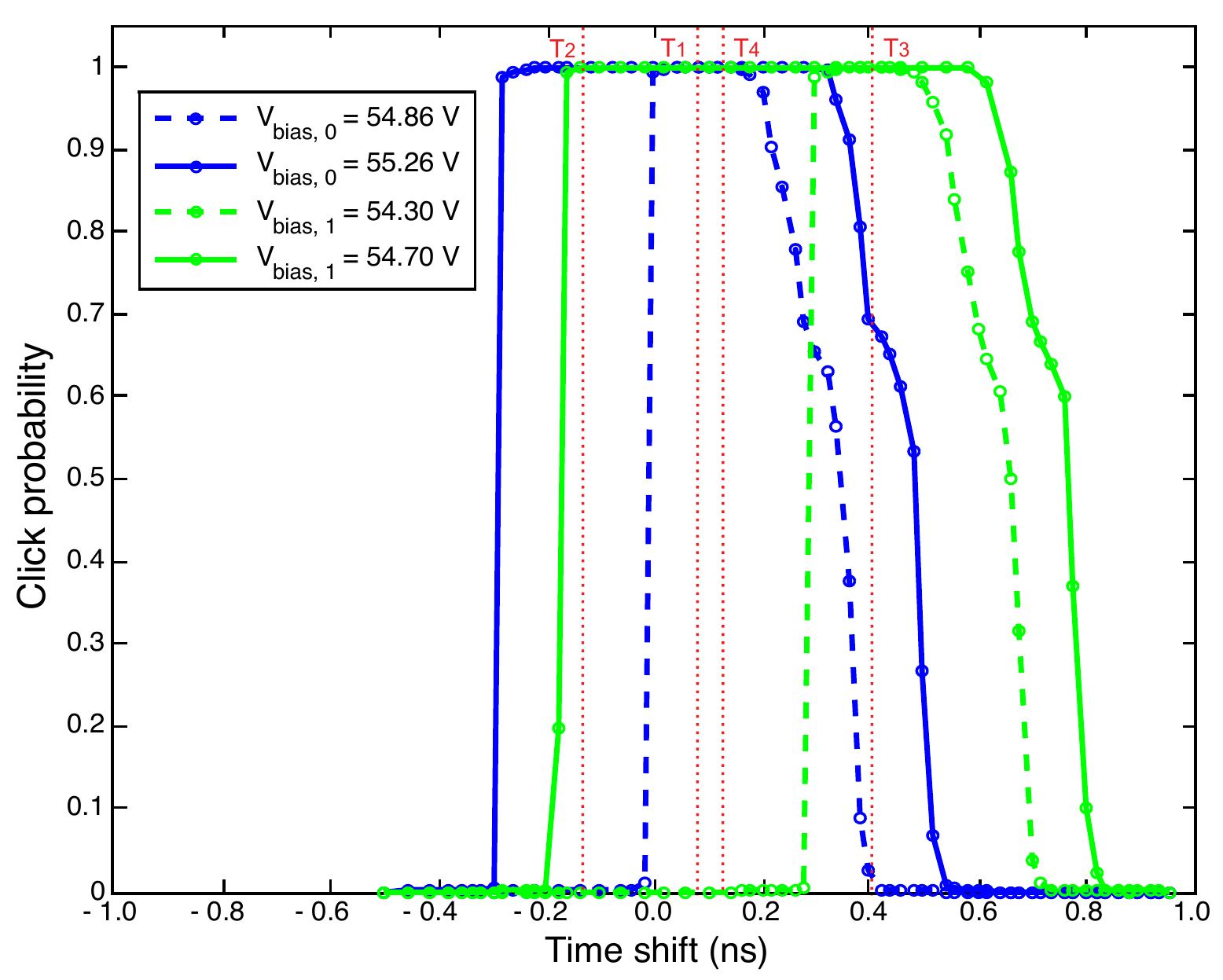}
	\caption{Click probabilities under blinding attack versus relative time shift of trigger pulse. Solid curves give the detection probability at the original $V_\text{bias}$, and dashed curves give the detection probability at lower $V_\text{bias}$. Note that the latter extends over a relatively narrower time window. The blinding power is  $0.38~\milli\watt$. The energy of trigger pulse for D0 is $0.22~\pico\joule$ and for D1 is $0.19~\pico\joule$. These energy levels are marked as red $\times$ in~\cref{fig:triggerresult}.}
	\label{fig:detection_efficiency_time}
\end{figure}

Second, we test the correlation between time shift of trigger pulse and click probability of blinded detector. The trigger pulse energy we use in this test for D1 is slightly lower than that of D0, but both levels of energy are above $E_\text{always,0,1}^\text{gate}$ in~\cref{fig:triggerresult} marked as red $\times$. The measurement result is shown in~\cref{fig:detection_efficiency_time}.

This testing result illustrates another method to attack the countermeasure: randomly adjusting the time shift of the trigger pulse. For D0, after fixing the suitable energy level of trigger pulse, Eve can always trigger a click by choosing time shift $T_1$, but only trigger a click at higher $V_\text{bias}$ by choosing $T_2$. Similarly, if target detector is D1, the detector always clicks at $T_3$, but only clicks at higher $V_\text{bias}$ at $T_4$. Then, when Eve sends trigger pulse to control D0, she randomly selects $T_1$ and $T_2$ with equal probability $P_0/2$ to reproduce the correct detection efficiencies of D0. Eve utilizes the same strategy for D1 to achieve correct detection probabilities, $P_1/2$ and $P_1$. In this way, Eve also hacks Clavis2 system tracelessly.

Generally, a finite set of decoy detection efficiency levels $\eta_1 < \eta_2 < \eta_3 < ... < \eta_n$ can be hacked by properly setting probabilities of different attacking energy levels or time-shifts. We take energy levels of trigger pulse as an example. According to the result in~\cref{fig:detection_efficiency_energy}, it is reasonable to extrapolate that we can find $n$ distinct levels of trigger pulse energy $E_1 > E_2 > E_3 > ... > E_n$ in this situation. Then Eve can apply $E_k$ ($k = 1, ..., n$) with probability $q_k$ to satisfy $\eta_k = \sum_{i=1}^k q_i$. This would reproduce every expected value of $\eta_k$ and hack the system. We have so far assumed that applying energy level $E_{k}$ causes zero click probability for decoy levels up to $\eta_{k-1}$, and 100\% click probability for $\eta_{k}$ and above. However this is not a necessary condition. More generally, under energy $E_{k}$, the click probability for efficiency level $\eta_{i}$ is $\beta ^{E_{k}} _{\eta_{i}}$. To reproduce the expected efficiencies, we need to satisfy the following set of equations:
\begin{align}
&q_1\beta_{\eta_1}^{E_1} + q_2\beta_{\eta_1}^{E_2} + ... + q_n\beta_{\eta_1}^{E_n} = \eta_1\nonumber\\
&q_1\beta_{\eta_2}^{E_1} + q_2\beta_{\eta_2}^{E_2} + ... + q_n\beta_{\eta_2}^{E_n} = \eta_2\nonumber\\
&......\nonumber\\
&q_1\beta_{\eta_n}^{E_1} + q_2\beta_{\eta_n}^{E_2} + ... + q_n\beta_{\eta_n}^{E_n} = \eta_n.
\end{align}
We might solve these equations to get values $0 \leq q_k < 1$. A worse case would be if Eve cannot find values of all $q_k$, which means she may only have a partial control of Bob's $\eta_k$. However, it still breaks the assumption in the security proof~\cite{lim2015} that Eve cannot form faked states with click probability conditional on Bob's randomly chosen efficiency. For quantitative analysis, an updated security proof would be needed first.

From the above testing and analysis of the implementation that changes $V_\text{bias}$, we can guess that an alternative implementation that changes $V_\text{gate}$~\cite{lim2015} or adds an intensity modulator in front of the detectors~\cite{moroder2009}, may leave a similar loophole. If we apply the intensity modulator, the energy of the trigger pulse arriving at the detector is not constant but depends on the modulation. However, this case is similar to gate voltage modulation, as we only consider the total energy from the gate signal and trigger pulse. Therefore, we will get similar results as~\cref{fig:detection_efficiency_energy,fig:detection_efficiency_time}, but the amount of trigger pulse energy and time shift might be different.

The reason for this practical loophole is a wrong assumption made by Lim and his colleagues~\cite{lim2015}. They assume Eve cannot generate faked states that trigger detections with probabilities that are {\em proportional} to the original photon detection efficiency. Here we have proved this is in fact possible. Therefore, the model of a practical detector should be more precise in security analysis, if one wishes to close the detector control loophole without resorting to measurement-device-independent QKD.

\section{Our attacks in a black-box setting}
\label{sec:attack-black-box}

According to Kerckhoffs' principle \cite{kerckhoffs1883}, Eve always knows everything about the algorithms and hardware of Alice's and Bob's boxes, including the precise values of equipment parameters. The classical security community practices Kerckhoffs' principle since 1970's, and widely agrees that this is a good approach to implementation security \cite{naor2003}. This is supported by many examples of cryptographic systems that did not follow this principle and were compromised \cite{singh1999}. The quantum academic community certainly agrees that QKD should be made secure in this setting, which is necessary for QKD being unconditionally secure~\cite{lo1999, shor2000, lutkenhaus2000, mayers2001, gottesman2004, renner2005}.

However, it is also a practically interesting question if any proposed attack can be mounted on today's commercial QKD systems in a black-box setting, when Eve only has access to the public communication lines but cannot directly measure signals and values of analog parameters inside Alice's and Bob's boxes \cite{gisin2015}. In this realistic scenario, Eve may purchase (or acquire by other means) a sample of the system hardware, open it, make internal measurements and rehearse her attacks on it. Then she has to eavesdrop on her actual target, an installed system sample in which she has not had physical access to the boxes. Although the latter sample can be of the same model and design, it will generally have different values of internal analog parameters, owing to sample-to-sample variation in system components. A full implementation of our attacks in this scenario remains to be tested. In this setting it will be of utmost importance for Eve to avoid triggering clicks in the absence of the gate, because this would very quickly brick the system and risk revealing her attack attempt. The original blinding attack that applies the trigger after the gate becomes very sensitive to precise values of thresholds in the presence of the first version of countermeasure (\cref{fig:originalattack}). For this reason we think the countermeasure will likely be triggered by the original attack in the realistic black-box setting.

Our modified attack that applies the trigger inside the gate will likely avoid triggering the alarm, because the no-gate threshold energies are much higher that the energies required for detector control (\cref{fig:triggerresult}). It also tolerates some fluctuation in experimental parameters for detector control. For example, when Eve applies $0.38~\milli\watt$ blinding power, $252~\femto\joule$ trigger pulse energy, and times her trigger pulse at the middle of the gate, we have verified that the attack still works perfectly for up to $\pm 21\%$ change in the trigger energy (see \cref{fig:triggerresult}) or up to $\pm 1.3~\nano\second$ change in the trigger timing. This makes it robust against reasonably expected fluctuations and imprecision of the system parameters. In particular, the timing accuracy required for our attack in much coarser than the several tens of picoseconds precision Alice and Bob use in normal operation \cite{jain2011}. The trigger energy setting precision is similar to the original attack that required $\pm 16\%$ \cite{lydersen2010a}.

Eve may need a few attempts to set a correct trigger energy when attacking a new copy of the system. She can do this by starting at a low trigger energy and attempting several increasing values of energy while watching the classical traffic Alice-Bob for the success or failure of the QKD session she has attacked \cite{makarov2005}. A QKD session that fails because of too low detection efficiency is a naturally occurring event that is part of normal system operation, does not raise an alarm and is recovered from automatically in Clavis2 \cite{jain2011, makarov2015}.

A full two-level implementation of the countermeasure may require Eve to run more attempts, because of a finer degree of control required over the trigger pulse energy and timing. Yet, similarly to the first countermeasure implementation, the no-gate trigger energy that would raise alarm remains safely well above the energies required for detector control. The practicality of attack in the black-box setting is thus difficult to predict without having the actual industrial implementation of the full countermeasure, and actually demonstrating the full attack, which can be a future study.

\section{Conclusion}
\label{sec:conclusion}

We have tested the first implementation of the countermeasure against the blinding attack in the commercial QKD system Clavis2. Our testing result demonstrates that presently implemented countermeasure is effective against the original blinding attack but not effective against a modified blinding attack. The modified attack fully controls Bob's single-photon detectors but does not trigger the security alarm. The modified attack is similar to the original detector blinding attack~\cite{lydersen2010a} with the only difference that the trigger pulses are time-aligned to coincide with the detector gates, instead of following it. We argue that this attack should be implementable in practice against an installed QKD communication line where Eve does not have physical access to characterising Alice and Bob, however such full demonstration has not yet been done, to our knowledge.

We have also tested the full proposed implementation of countermeasure with two non-zero efficiency levels, and found its security to be unreliable despite predictions of the theory proposal~\cite{lim2015}. From the current testing results, bright-pulse triggering probabilities of the blinded detectors depend on several factors including $V_\text{bias}$, timing and energy of the trigger pulse (see \cref{sec:discussion-of-full-countermeasure}). This in principle allows Eve to compromise the full countermeasure implementation.

We have tested the countermeasure implemented with the gated single-photon detectors (SPDs). The idea of random detection efficiency can be applied to other types of SPDs that are also sensitive to the blinding attack: free-running SPDs~\cite{sauge2011} and superconducting nanowire SPDs~\cite{lydersen2011c}. However, the countermeasure based on these detectors might still be hackable. Since the efficiencies of these types of SPDs depend on the bias voltage or current, varying these bias signals likely changes other parameters inside the SPD and its electronics. Therefore, when we randomize the detection efficiency, other degrees of freedom might be changed as well. Eve has a chance to exploit these side channels to hack the countermeasure. Of course, the exact outcome cannot be known until the countermeasures in different types of detectors are experimentally tested.

According to our testing result, this countermeasure is not as reliable as would be expected in a high-security environment of QKD. Although an ideal industrial countermeasure has not been achieved, everybody now has a more clear concept about the detector loopholes. This procedure emphasizes the necessity of security testing every time practical QKD systems are developed or updated. We only can reach the final practical security of any QKD system after several iterations of implementation development and testing verification. Our countermeasure testing also illustrates that patching a loophole is still time-consuming and difficult. However, addressing practical vulnerabilities at the design stage of a QKD system is both cheaper and less messy than trying to retrofit patches on an existing deployed solution. Addressing security at the design stage should be the goal whenever possible.

\section*{Acknowledgment}
We thank C.~C.~W.~Lim, N.~Gisin, and E.~Anisimova for discussions. This work was supported by Industry Canada, NSERC (programs Discovery and CryptoWorks21), CFI, Ontario MRI, US Office of Naval Research, ID~Quantique, European Commission's FET
QICT SIQS and EMPIR 14IND05 MIQC2 projects. P.C.\ was supported from Thai DPST scholarship.

\appendix

\section{Background}
\label{sec:background}

\begin{figure}[b]
	\includegraphics[width=1\columnwidth]{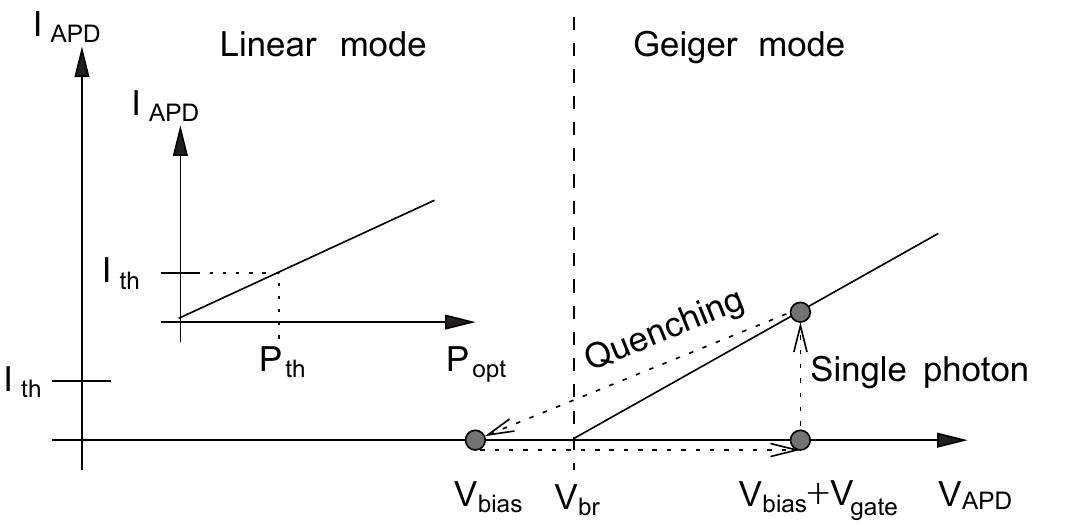}
	\caption{Linear-mode and Geiger-mode APD operation (reprinted from~\cite{lydersen2010a}).}
	\label{fig:modes}
\end{figure}

In this section, we recap the operating principle of the single-photon detector, its implementation in Clavis2, and the original blinding attack~\cite{lydersen2010a}. Most available single-photon detectors are APDs operating in Geiger mode, in which they are sensitive to single photons~\cite{cova2004}. As shown in \cref{fig:modes}, when the APD is reverse-biased above its breakdown voltage $V_\text{br}$, a single photon can cause a large current $I_\text{APD}$. If this current exceeds the threshold $I_\text{th}$, electronics registers this as a photon detection (a `click'). After that, an external circuit quenches the avalanche by lowering the bias voltage $V_\text{APD}$ below $V_\text{br}$, and the APD comes into a linear mode. If the APD is illuminated by bright light (which does not happen in normal single-photon operation but can happen during an eavesdropping attack), $I_\text{APD}$ in the linear mode is proportional to the incident bright optical power $P_\text{opt}$. $I_\text{th}$ then becomes a threshold on the incident optical power $P_\text{th}$ that makes a click.

\begin{figure}
	\includegraphics[width=\columnwidth]{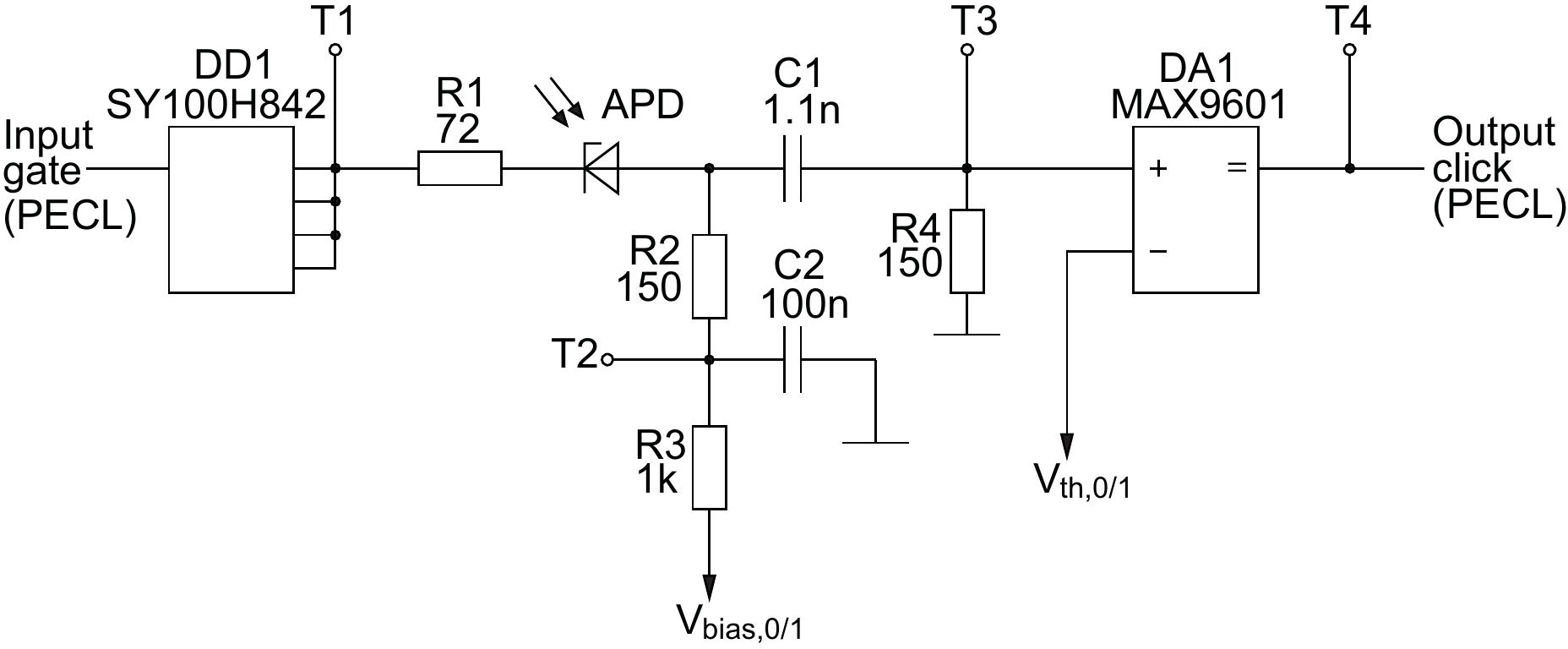}
	\caption{Equivalent detector bias and comparator circuit, as implemented in Clavis2 (reprinted from \cite{lydersen2010a}).}
	\label{fig:circuit}
\end{figure}

From an engineering view, the detector can be analyzed by its circuit. \Cref{fig:circuit} shows an equivalent circuit diagram of the two detectors used in Clavis2. When no gate signal is applied, the APDs are biased slightly below their $V_\text{br}$ by the negative high-voltage supply $V_\text{bias,0} = -55.26~\volt$, $V_\text{bias,1} = -54.70~\volt$ \footnote{Using values from the sample of Clavis2 tested in our present study at the University of Waterloo, which is a different sample than in Refs.~\onlinecite{lydersen2010a, lydersen2010b, wiechers2011}}. To bring the APD into Geiger mode, an additional $3~\volt$ high, $2.8~\nano\second$ long pulse is applied through a logic level converter DD1. The anode of the APD is AC-coupled to a fast comparator DA1. Since the capacitor C1 blocks the DC component, only when the current flowing through the APD changes, it generates a pulse as the input of DA1. If the peak voltage of this pulse is greater than the positive threshold $V_\text{th} = 70~\milli\volt$, the comparator produces a logic output signal indicating a click. Once a click in either of the two Bob's detectors is registered, the next 50 gates will not be applied to both detectors, which constitutes a deadtime to reduce afterpulsing.

If Eve sends a bright c.w.\ illumination to the gated detectors, the bright light makes the APD generate a significant photocurrent that monotonically increases with the optical power $P_\text{opt}$. When we consider effects of this current on the whole detector circuit~(\cref{fig:circuit}), the most useful one is a reduction of the voltage across the APD $V_\text{APD}$. Although the high-voltage supply $V_\text{bias}$ stays constant, the photocurrent causes a significant voltage across $\text{R3} = 1~\kilo\ohm$, thus $V_\text{APD}$ drops. If we apply enough illumination power, $V_\text{APD}$ will be less than $V_\text{br}$ even inside the gate, and the APD then always stays in the linear mode. The detector becomes blind to single photons. In our testing, we measure the voltage at test point T2 $V_\text{T2}$ in~\cref{fig:circuit} and refer to this voltage as $V_\text{APD}$ in the text. $V_\text{T2}$ is close to real $V_\text{APD}$, because $\text{R1}+\text{R2} \ll R3$ [precisely, $V_\text{APD} = V_\text{T2} + (V_\text{T2}-V_\text{bias})(\text{R1}+\text{R2})/\text{R3}$].

After blinding Bob's detectors, Eve can conduct a faked-state attack. Eve first intercepts all photons sent by Alice. Whenever Eve detects a photon, she sends the same state to Bob via a bright trigger pulse of a certain energy, superimposed on her blinding illumination. Only if Bob chooses the same measurement basis as Eve and applies the gate, one of Bob's detectors will click and he will get the same bit value as Eve. Otherwise, there is no click at Bob's side. During the sifting procedure, Alice and Bob keep the bit values when they have chosen the same basis, and so does Eve. Therefore Eve has identical bit values with Bob, introduces no extra QBER, and does not increase the alarm counter. Eve then listens to the public communication between Alice and Bob and performs the same error correction and privacy amplification procedures as them, to obtain an identical copy of their secret key~\cite{lydersen2010a}.

\section{Analysis of processes in the detector}
\label{sec:understanding-apd-operation}

\begin{figure}
	\includegraphics[width=\columnwidth]{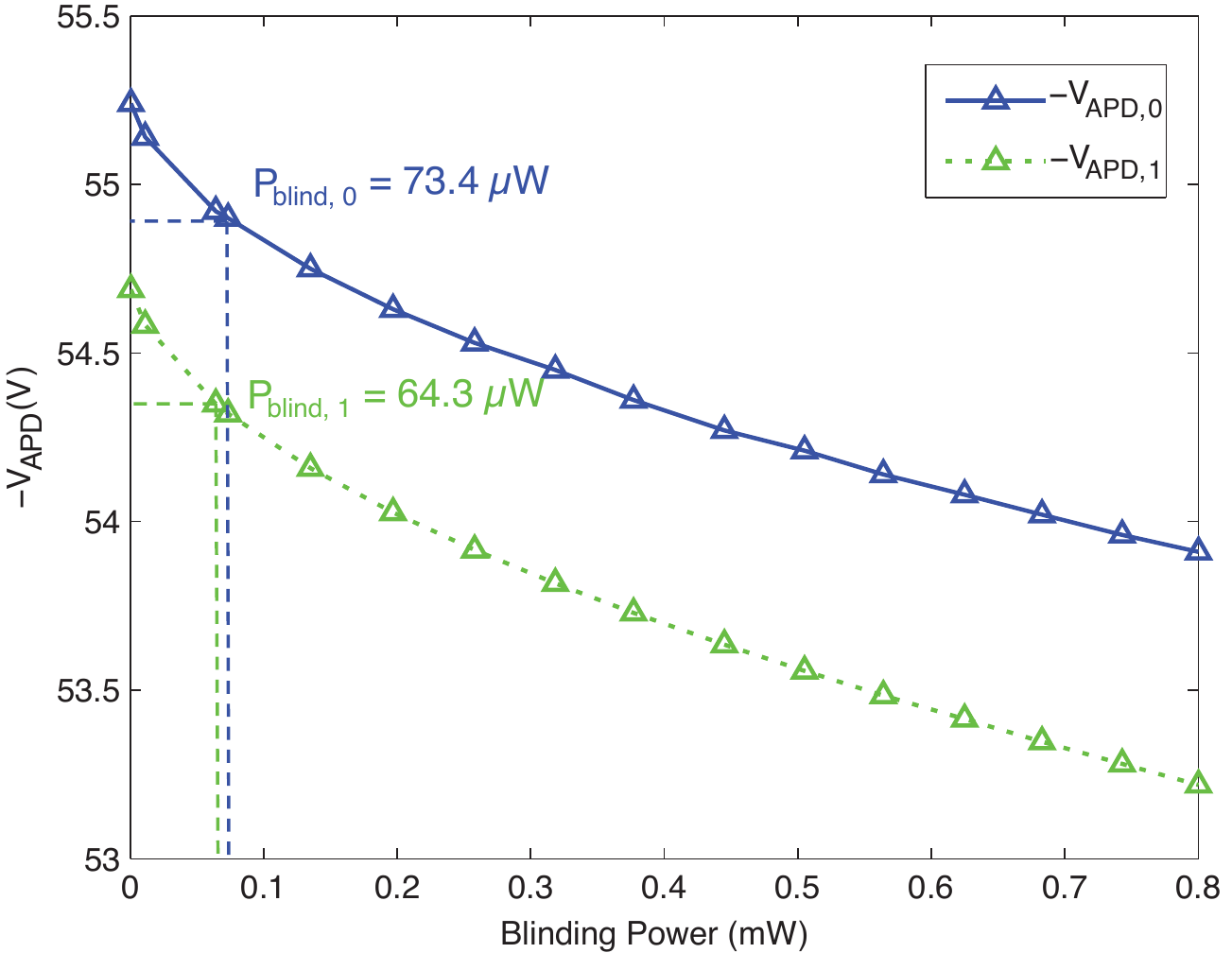}
	\caption{Bias voltage of APDs versus c.w.\ blinding power.}
	\label{fig:biasvoltage}
\end{figure}

For further understanding of the detector behaviour under successful blinding attack, we attempt to quantitatively model electrical and thermal processes in it. As we mentioned previously, the bias voltage decreases when the blinding power is applied. A measured relationship between $V_\text{APD}$ and continuous blinding power is shown in~\cref{fig:biasvoltage}. Detector 0 is blinded at $P_\text{opt} > P_\text{blind,0} = 73.4~\micro\watt$ and detector 1 is blinded at $P_\text{opt} > P_\text{blind,1} = 64.3~\micro\watt$. Higher blinding illumination leads to lower bias voltage. This is consistent with the same measurement done for the original blinding attack~\cite{lydersen2010a}.

In a detector blinded by c.w.\ laser illumination, the gain factor is affected by not only the power of blinding laser, but also the gate signal. When the APD is blinded and forced to work in the linear mode, it can be treated as an ordinary photodiode with a finite internal gain. Photoelectrons and holes are accelerated by a high electric field and initiate a chain of impact ionizations that generates secondary electron-hole pairs. Thus, the APD has an internal multiplication gain factor $M > 1$, since one photon can yield many electrons of photocurrent flowing in the circuit. When $V_\text{APD}$ is much lower than $V_\text{br}$, $M$ will be close to 1. However, the APD may not have any significant photosensitivity below so-called punch-through voltage, below which the electrical field does not extend into the absorption layer of InGaAs/InP heterostructure~\cite{hiskett2000}.

\begin{figure}
	\includegraphics[width=1\columnwidth]{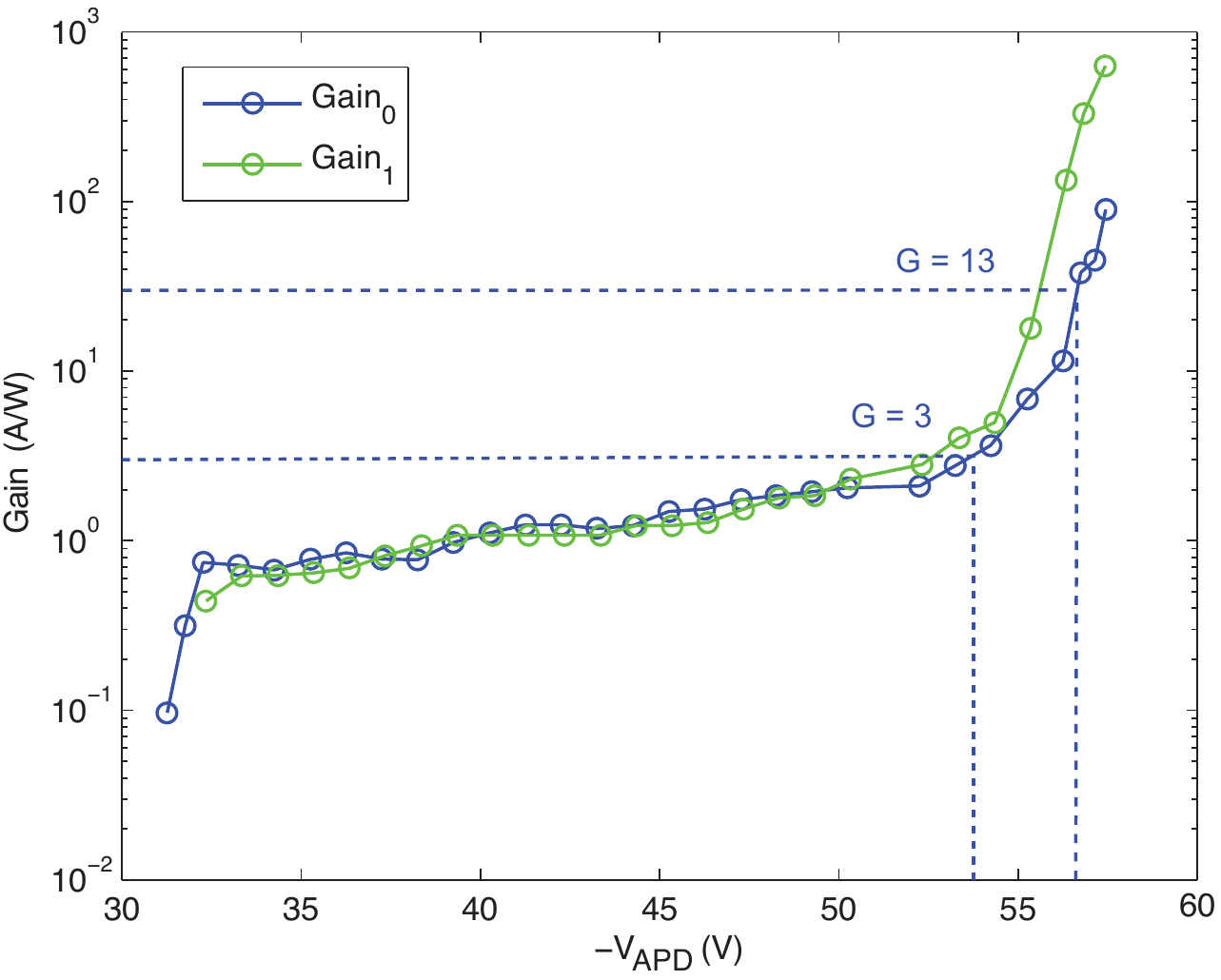}
	\caption{Gain versus APD bias voltage. Values of gain for bias voltages below $31~\volt$ were negligibly low for a practical attack, and below the sensitivity of our measurement method.}
	\label{fig:gain}
\end{figure}

We have done a measurement of small-signal gain $G$ of the APDs in Clavis2 by measuring their photocurrent response to a short optical pulse input. The results are shown in~\cref{fig:gain}. There is virtually no photosensitivity below the punch-through voltage of about $31~\volt$. Above that voltage $G$ starts at $\sim 0.7~\ampere\per\watt$ (corresponding to $\sim 60\%$ quantum efficiency assuming $M = 1$), then rises above $100~\ampere\per\watt$ closer to $V_\text{br}$. The gain values measured at $V_\text{br} - 2~\volt$ are $\sim 7$ and $\sim 10~\ampere\per\watt$, which is consistent with values from data sheets of commercial APDs. From the above measurements, we know that Eve can vary the amount of blinding power to the detectors to control the bias voltage and thus the gain factor.

After we blind Bob's detectors in Clavis2, the gain factor is greater during the $2.8~\nano\second$ gate duration, because the gate signal raises $V_\text{APD}$. Thus the electrical charge generated by the APD in response to a trigger pulse applied in the gate is greater than when it's applied outside the gate. For example, in~\cref{fig:gatetriggersingal}(c), the gate pulse alone contributes $1.053~\pico\coulomb$ extra charge on top of the current that would be generated without the gate. When the trigger pulse is applied after the gate [\cref{fig:gatetriggersingal}(d)], the total charge of the two pulses is $1.467~\pico\coulomb$; however, when the trigger pulse is moved into the gate [\cref{fig:gatetriggersingal}(e)], the total charge rises to $1.613~\pico\coulomb$. Therefore, a greater gain factor during the gated time helps the pulse to cross  the threshold.

We have attempted to model the increased gain due to the gate. In our model, we consider a thermal effect and an internal resistance of the APD. On the one hand, an increased temperature raises $V_\text{br}$~\cite{sze2007}. Electrical heating ($V_\text{APD}\cdot I_\text{APD}$) and the absorption of the blinding power result in a heat dissipation: $61.2~\milli\watt$ for detector 0 and $66.03~\milli\watt$ for detector 1~\footnote{Under $0.564~\milli\watt$ blinding power, $V_\text{APD,0} = 54.14~\volt$, $I_\text{APD,0} = 1.12~\milli\ampere$. Heat dissipation of detector 0: $54.14~\volt\cdot 1.12~\milli\ampere + 0.564~\milli\watt = 61.2~\milli\watt$; $V_\text{APD,1} = 53.484~\volt$, $I_\text{APD,1} = 1.224~\milli\ampere$, Heat dissipation of detector 1: $53.484~\volt\cdot 1.224~\milli\ampere + 0.564~\milli\watt = 66.03~\milli\watt$.}. Then, an estimated $190~\kelvin/\watt$ thermal resistance~\cite{lydersen2010b} between each APD chip and the cold plate converts the power dissipation into the increased temperature. The temperature-dependent breakdown voltage increases with the coefficient of about $0.1~\volt/\kelvin$~\cite{lydersen2010b}. As a result, $V_\text{br}$ increases by $1.16~\volt$ ($1.25~\volt$) for detector~0 (1). \Cref{fig:gain} shows the relation between gain factor and the actual $V_\text{APD}$ in the linear mode. When $V_\text{APD}$ is close to $V_\text{br}$, the gain factor increases rapidly. On the other hand, we suppose the APD has a passive internal resistance, so the internal bias voltage across the ideal photodiode is less than the value of $V_\text{APD}$ we test. By measuring the voltage of a stable avalanche pulse and calculating the current trough the detector circuit when avalanche happens, we obtain the internal resistance of $330~\ohm$ in detector 0 and $275~\ohm$ in detector 1. Therefore, the real bias voltage under blinding attack shown in~\cref{fig:gatetriggersingal}(c--f) is $53.77~\volt$, which corresponds to $G = 3~\ampere/\watt$ in detector~0 as shown in~\cref{fig:gain}. When $3~\volt$ gate is applied, the bias voltage becomes $56.77~\volt$ which corresponds to $G = 13~\ampere/\watt$ in~\cref{fig:gain}. However, the measured charges in~\cref{fig:gatetriggersingal}(d) and (e) illustrate much less gain change: $G = 1.3~\ampere/\watt$ at $53.77~\volt$ and $G = 1.76~\ampere/\watt$ at $56.77~\volt$~\footnote{When we apply a $0.32~\pico\joule$ trigger pulse after the gate, this single trigger pulse contributes $0.414~\pico\coulomb$ charge which is the difference between the total charges in~\cref{fig:gatetriggersingal}(c) and (d). $G = 0.414~\pico\coulomb/ 0.32~\pico\joule = 1.3~\ampere/\watt$. When we apply a $0.32~\pico\joule$ trigger pulse during the gate, this single trigger pulse contributes $0.56~\pico\coulomb$ charge which is the difference between the total charges in~\cref{fig:gatetriggersingal}(c) and (e). $G = 0.56~\pico\coulomb/ 0.32~\pico\joule = 1.76~\ampere/\watt$.}. The discrepancy may be explained by a larger actual thermal resistance between the APD and the cold plate than we estimate, which should be verified in future research.

\def\bibsection{\medskip\begin{center}\rule{0.5\columnwidth}{.8pt}\end{center}\medskip} 


\bibliography{bibtex_library}

\end{document}